\documentclass[pdftex,dvipsnames,usenames,aps,superscriptaddress,twocolumn,amsmath,showkeys]{revtex4-2}

\usepackage{hyperref}
\usepackage{float}
\usepackage{dcolumn}
\usepackage{graphicx}
\usepackage{color}
\usepackage{amssymb}
\usepackage{bold-extra}
\usepackage{caption}
\usepackage{subcaption}
\usepackage{verbatim}
\usepackage[utf8]{inputenc}
\usepackage{multirow}
\usepackage{soul}

\newcolumntype{d}[1]{D{.}{.}{#1} }
\newcolumntype{h}[1]{D{-}{-}{#1} }

\def\vB{{\bf{v}\times\bf{B}}}
\def\vvB{{\bf{v}\times\left(\bf{v}\times\bf{B}\right)}}

\def\beq{\begin{equation}}
\def\eeq{\end{equation} }
\def\bea{\begin{eqnarray}}
\def\eea{\end{eqnarray}}

\def\figref#1{Fig.~\ref{fig:#1}}
\def\figlab#1{\label{fig:#1}}  
\def\eqref#1{Eq.~(\ref{eq:#1})}

\def\eqlab#1{\label{eq:#1}}
\def\tabref#1{Table~\ref{tab:#1}}
\def\tablab#1{\label{tab:#1}}  
\newcommand*{\secref}[1]{Section~\ref{sec:#1}}
\newcommand*{\seclab}[1]{\label{sec:#1}}
\newcommand{\Omit}[1]{}

\def\Xmax{X_{\rm max}}
\def\Xrh{X_{\rm z}}
\def\Prog#1{} 

\def\RUG{Kapteyn Astronomical Institute, University of Groningen, P.O. Box 72, 9700 AB Groningen, Netherlands}
\def\AIVUB{Astrophysical Institute, Vrije Universiteit Brussel, Pleinlaan 2, 1050 Brussels, Belgium}
\def\VUB{Interuniversity Institute for High-Energy, Vrije Universiteit Brussel, Pleinlaan 2, 1050 Brussels, Belgium}

\def\IMAPP{Department of Astrophysics/IMAPP, Radboud University Nijmegen, Nijmegen, The Netherlands}

\def\SoE{Physics Education Department, School of Education, Can Tho University, Campus II, 3/2 Street, Ninh Kieu District, Can Tho City, Viet Nam}
\def\CVL{Department of Education and Training, 3/2 Street, Ninh Kieu District, Can Tho City, Viet Nam}
\def\SoS{Department of Physics, Can Tho University, 3/2 Street, Ninh Kieu, Can Tho City 94000, Vietnam}
\def\FoP{Faculty of Physics, University of Warsaw - Warsaw, Poland}

\begin{document}

\title{Determining Atmospheric Electric Fields using MGMR3D}

\author{T.~N.~G.~Trinh} \email[]{ttngia@ctu.edu.vn}  \affiliation{\SoE}
\author{O.~Scholten}  \email[]{o.scholten@rug.nl } \affiliation{\RUG}   \affiliation{\VUB}
\author{S.~Buitink} \affiliation{\AIVUB} \affiliation{\IMAPP}
\author{K. D.~de~Vries} \affiliation{\AIVUB} 
\author{P.~Mitra} \affiliation{\AIVUB}\affiliation{\FoP}
\author{T.~Phong~Nguyen} \affiliation{\SoS}
\author{D. T.~Si} \affiliation{\CVL}


\begin{abstract}
Cosmic-ray particles impinging on the atmosphere induce high-energy particle cascades in air, an Extensive Air Shower (EAS), emitting coherent radio emission. This emission is affected by the presence of strong electric fields during thunderstorm conditions. To reconstruct the atmospheric electric field from the measured radio footprint of the EAS we use an analytic model for the calculation of the radio emission, MGMR3D. In this work we make an extensive comparison between the results of a microscopic model for radio emission, CoREAS, to obtain an improved parametrization for MGMR3D in the presence of atmospheric electric fields, as well as confidence intervals. The approach to extract the electric field structure is applied successfully to an event with a complicated radio footprint measured by LOFAR during thunderstorm conditions. This shows that, with the improved parametrization, MGMR3D can be used to extract the structure of the atmospheric electric field.
\end{abstract}


\maketitle

\section{Introduction}
Lightning is a common phenomena but the detailed understanding of its generation and development is still unknown. One of the reasons why this topic is still under investigation is that atmospheric electric fields inside thunderclouds are difficult to measure. Existing measurements, from balloons and airplanes, are limited because the measurements depend on the paths of the balloons or the aircrafts will affect the nature of the thunderclouds. A new method to determine atmospheric electric fields is using their influence on radio emission emitted from extensive air showers~\cite{Schellart:2015}. Unlike the balloon and airplane measurements, this unique tool is not limited by the wind conditions and is sensitive to a large part of the cloud.

When a cosmic ray enters the atmosphere of the earth, it interacts with air molecules and generates a particle cascade, called an extensive air shower (EAS). The particles in the EAS move with velocities near the speed of light and are concentrated in the thin shower front. In an EAS during fair-weather conditions, called fair-weather showers, the electrons and positrons are deflected in opposite directions by the Lorentz force caused by the geomagnetic field. This induces an electric current in the shower front that is transverse to the shower axis. Since the electric current changes as a function of height, due to the change of the number of particles in the EAS, radio waves are emitted~\cite{Scholten:2008, Werner:2012, Scholten:2009}.
A secondary, yet important, contribution to the radio emission is caused by an excess of electrons in the shower front due to knock-out from atmospheric molecules by shower particles. This process creates a radio pulse that is linearly polarized but oriented radially to the shower axis~\cite{Askaryan:1962, Krijn:2010}. This charge excess emission interferes with the geomagnetic emission that is linearly polarized along the direction of the Lorentz force.

An EAS that occurs during thunderstorm conditions when there are strong atmospheric electric fields, is called a thunderstorm shower, and produces radio emission that differs considerably from that of fair weather showers~\cite{Schellart:2015, Trinh:2016, Trinh:2020}.
In thunderstorm showers, the atmospheric electric field exerts an electric force on the electrons and positron that is usually much stronger than the Lorentz force and this affects the radio emission. The component of the electric field perpendicular to the shower axis will change the direction and magnitude of the transverse current changing the shape of the radio footprint.
The electric field parallel to the shower axis accelerates the electrons or positrons, depending on its direction, and thus they gain more energy. As a result, these particles will generate additional low-energy particles. However, these additional particles travel with a velocity smaller than the speed of light and thus they trail far behind the shower front.
Thus, their radiation is not added coherently in the frequency range from 30~MHz to 80~MHz at which typical cosmic-ray airshower radio detectors operate~\cite{Trinh:2016}.

There are several models to describe radio emission from air showers. Microscopic models such as ZHAires~\cite{Alvare:2012} and CoREAS~\cite{Huege:2013} are based on full Monte-Carlo simulation codes. Macroscopic models such as MGMR~\cite{Scholten:2008}, EVA~\cite{Werner:2012} and Radio Morphing~\cite{Zilles:2020} calculate the emission of the bulk of electrons and positrons described as currents. SELFAS2~\cite{Marin:2012} is a mix of macroscopic and microscopic approaches. Recently, MGMR3D~\cite{Scholten:2018} has been introduced that uses a multi-dimensional parametrization of the current density in the air showers to calculate the radiation field using Maxwell equations.

Because atmospheric electric fields influence the radio emission from air showers, the radio footprint, as is measured on the ground, can be used to determine the strength, direction, and the altitude dependence of the atmospheric electric field along the path of the air showers~\cite{Schellart:2014}.
Since we have only a forward model, i.e.\ calculate the footprint by assuming an electric field configuration, we have to apply a search algorithm to extract the field structure. This only works with a fast model calculation that is deterministic, i.e.\ non Monte-Carlo based.
MGMR3D is potentially such a model. However, MGMR3D relies on a parameterization of the emitting currents. To gauge these parameters we perform an extensive comparison between MGMR3D and the full Monte-Carlo code CoREAS.

In Ref.~\cite{Scholten:2018,Trinh:2020} a parametrization was proposed based on CONEX-MC~\cite{Bossard:2001} simulations. In particular it had been assumed that the induced drift velocity, driving the currents, is inversely proportional to the air density, $v_d\propto 1/ \rho$.
This assumes that the drift velocity is inversely proportional to the friction force, the collision frequency.
In Ref.~\cite{Mitra:PhD} it was shown that an inverse dependence of the drift velocity on $v_d\propto 1/ \sqrt{\rho}$ yields a much better agreement with the results of microscopic calculations.
The $1/\sqrt{\rho}$ dependence is characteristic of an equilibrium situation, terminal velocity, where the acceleration of the fast electron by the Lorentz force is compensated by the average friction due to collisions with air molecules~\cite{Mitra:PhD}. This has led to a considerable improved parametrization for the description of fair weather showers in MGMR3D~\cite{Mitra:PhD}.
This different proportionality for the drift velocity has made us to revisit the parametrizations that enter in MGMR3D in the presence of atmospheric electric fields, by making an extensive comparison with a large number of CoREAS calculations for random shower directions and various structures of the atmospheric electric fields. The improved parametrization allows us to extract the fields for more complicated configurations that were inaccessible before.

In \secref{coreas} we present a very short review of the most important aspects of CoREAS that describes radio emission on a microscopic level and which we consider as `the truth' to which we gauge the parameters that enter in the modeling of the currents entering in MGMR3D, as discussed in \secref{mgmr}. We first review the essentials of the parametrization for fair-weather showers~\cite{Mitra:PhD} after which we present the discussion of the parametrization in the presence of electric fields.
In \secref{appl} we apply the improved model to data for the event which was not possible to analyze in Ref.~\cite{Trinh:2020} to extract the atmospheric electric field.

\section{CoREAS simulations}\seclab{coreas}

To gauge the parameters in the MGMR3D model, discussed in \secref{mgmr}, we perform extensive comparisons with the radio emission footprints from EAS as calculated from CoREAS, which we regard as the Monte-Carlo truth. CoREAS is a plug-in for the particle simulation code CORSIKA~\cite{Heck:1998} and is based on a microscopic description of the radiation mechanism.  In CoREAS the radio emission from each electron and positron  in  the shower is calculated, without making any assumptions on the type of radiation. In other words, CoREAS does not simulate the transverse-current and charge-excess components separately, but it produces the complete radiation field from the emissions of the individual particles. Therefore, CoREAS simulations are compute intensive with running times of the order of days. Using the electric field EFIELD option~\cite{Buitink:2007} in CORSIKA it is possible to simulate air showers passing through strong electric-field regions. For this work the radiation profile is calculated in the shower plane (the plane perpendicular to the shower axis passing through the point where the shower axis touches ground), with the $\hat{x}$-axis in the direction of $\vB$, along the direction of the Lorentz force, and $\hat{y}$-axis along $\vvB$. Here $\bold{v}$ is the velocity of the shower and $\bold{B}$ is the geomagnetic field. We use the shower plane instead of the ground plane since it allows for an easier interpretation. To construct a two-dimensional radio footprint, we run simulations for 160 antenna positions in a star-shaped pattern in the shower plane with eight arms, where each arm contains 20 antennas with a spacing of 25 m~\cite{Buitink:2013}.

To compare with the results from MGMR3D, we extract the values for the Stokes parameters at the 160 antenna positions and thus capture both intensity and polarization of the radio emission. Because the aim of this work is to develop a realistic approach to extract  electric fields from LOFAR~\cite{Haarlem:2013} data, we construct the Stokes parameters from the CoREAS simulations in the frequency band from 30~MHz to 80~MHz.
For each position in each simulated shower, the Stokes parameters are calculated as
\beq
I = \frac{1}{n}\sum_{i=0}^{n-1}\left ( \left | \varepsilon_{i,\vB} \right |^{2} +\left | \varepsilon_{i,\bf{v}\times \left(\vB\right)} \right |^{2}\right ),
\eeq
\beq
Q= \frac{1}{n}\sum_{i=0}^{n-1}\left ( \left | \varepsilon_{i,\vB} \right |^{2} -\left | \varepsilon_{i,\bf{v}\times \left(\vB\right)} \right |^{2}\right ),
\eeq
\beq
U+iV= \frac{2}{n}\sum_{i=0}^{n-1}\left(\varepsilon_{i,\vB} \varepsilon^*_{i,\bf{v}\times \left(\vB \right)} \right).
\label{Stokes}
\eeq
$\varepsilon_{i,x}$ is the complex-valued signal radiation field along direction $x$, where $i$ denotes the sample number and $x=\vB$ or $x=\bf{v}\times \left(\vB\right)$. $n$ is the number of time samples centered around the pulse-peak position.
Stokes $I$ is the intensity of the radio emission. Stokes $Q$ and $U$ are used to derive the linear polarization angle
\beq
\psi =\frac{1}{2}\mathrm{tan}^{-1}\left ( \frac{U}{Q} \right ),
\eeq
and Stokes $V$ is the intensity of the circularly polarized fraction of the radiation.

In order to perform chi-square fitting with realistic sensitivities, error bars (confidence levels) were assigned to the values of the Stokes parameters for the CoREAS simulated showers, given by
\beq
\sigma_{k,\left(I,Q,U, \mathrm{or}\,V\right)} =2\sqrt{\frac{\xi_I}{n}(I_k+\xi_I)}, \eqlab{err}
\eeq
where $n$ is the number of time samples as enters in the equations for the Stokes parameters Eq.~\ref{Stokes}, $I_k$ is the value of Stokes $I$ for a test antenna at position $k$ and $\xi_I$ is taken at the level of the instrumental noise measured in LOFAR antennas~\cite{Buitink:2013}. This definition of the error bars of the Stokes parameter guarantees that the relative error bars at positions near the core, where the intensity is usually large, is small while at large distances where the intensity is small, the relative error bars become large. 

\section{Parameterization in MGMR3D}\seclab{mgmr}

In MGMR3D~\cite{Scholten:2018}, the Maxwell's equations are solved in the far-field approximation to obtain the radio footprint, using a parametrized charge-current density that depends on atmospheric depth, radial distance from the shower axis, and distance from the shower front.
In earlier work,  presented in Ref.~\cite{Trinh:2020}, the parametrization of the charge-current density was obtained through a comparison of selected cases with the results of CONEX-MC and CoREAS calculations. The resulting parametrization was adequate for extracting the structure of relatively simple atmospheric electric fields, however for a thunderstorm EAS with a complicated radio footprints there were are large discrepancies  between the MGMR3D and CoREAS results~\cite{Trinh:2020}.

Following the approach presented in Ref.~\cite{Mitra:PhD} we tune the parametrization in MGMR3D by making a statistical comparison of the calculated radio footprint with the results of microscopic calculations for a large numbers of showers passing through different electric-field structures. The advantage of such an approach is that a comparison is made at the level of the important observables for later applications. In addition this approach allows to asses the intrinsic accuracy of obtaining the structure of the atmospheric electric field.

Since an extensive discussion of the parametrization for fair-weather showers is presented in Ref.~\cite{Mitra:PhD}, we will in this paper only summarize the essentials of the parametrization for fair-weather showers in \secref{FairWeather}. The results~\cite{Mitra:PhD} show that a good agreement between MGMR3D and CoREAS is obtained for fair-weather showers. The parametrization of the charge-current cloud in the presence of electric fields is discussed in \secref{Thunder}.

\subsection{Parameterization for fair-weather showers}\seclab{FairWeather}

Under fair-weather conditions only the Lorentz force, $\vec{F}_{\perp} = e\vec{v} \times \vec{B}$,  is acting on the charged particles in the EAS, causing the electrons and positrons to drift in opposite directions and thus induce an electric current. As presented in~\cite{Mitra:PhD, TerminalVelocity} the induced drift velocity depends on the square-root of the air density and can be parametrized as
\beq
\vec{v}_{\perp}(h)=\vec{F}_{\perp}/F_0 \frac{\sqrt{\rho(\Xmax)/\rho(h)}}{((\Xmax-X_v)/(X_z(h)-X_v) + 3.)} \;,
\eqlab{vh}
\eeq
with $X_v=50$~g/cm$^2$ and $F_0=62.5$~keV/m.
This -on first sight- surprising dependence on $\sqrt{\rho}$ is due to the fact that at the equilibrium sideways `friction' force, proportional to $\rho v_\perp^2$, equals the transverse component of the Lorentz force.
Since only the Lorentz force is acting, the drift velocity $\vec{u}_\perp$ is still non-relativistic and it is parameterized as
\beq
\vec{u}_\perp(h)=c\vec{v}_{\perp}(h)/\sqrt{1+v_{\perp}^2/u_0^2} \;,
\eqlab{drift-v}
\eeq
where the parameter $u_0$ is taken much larger than typical fair-weather drift velocities, resulting in a linear scaling of the drift velocity with the Lorentz force.

In MGMR3D the number of charged particles at a radial position $r_s$ from the shower axis and a longitudinal position $d_h$ from the shower front is written as $N f(d_h) w/r_s$, with $w(r_s)$ the normalized radial particle distribution, and $f(d_h)$ the normalized longitudinal particle distribution. The current is the product of the charged particle number and the drift velocity, where both depend on the penetration depth in the atmosphere, $X_z(h)$. The current is thus expressed as
\beq
\vec{J}_\perp(t_s,x_s,y_s,h)={w(r_s) \over r_s} \, f(d_h,r_s) \, N_{c}(X_z)\vec{u}_\perp(h) \;,
\label{current}
\eeq
at a radial distance $r_s=\sqrt{x_s^2+y_s^2}$ from the shower axis. The parametrization of the longitudinal shower profile for the current in the shower front is based on the Gaisser-Hillas formula~\cite{Gaisser:1977} for the dependence of the number of charged particles on $X_z$,
\beq
N_{c}(\Xrh)=  \left({ \Xrh-X_0 \over \Xmax -X_0}\right)^{\Xmax -X_0 \over \gamma} e^{\Xmax - \Xrh \over \gamma}
\label{Def-Nc-GH}
\eeq
where $\gamma$ is a parameter controlling the width of the distribution and $X_0$ the first-interaction point. $\Xmax$ is the atmospheric depth where the number of secondary particles reaches a maximum in the air shower.

The spatial distribution of the charged particle density is parametrized as
\beq
w(r_s)=N_w\,\xi (\xi+1)^{-2.5} \;; f(d_h,r_s)=N_f\, {\eta \over e^{\sqrt{\eta}} + 1}\;,
\eqlab{Def-w}
\eeq
with $\eta={d_h/ \lambda}$ where $d_h$ is the distance from the shower front, and  $\xi=r_s/R_0$ is a scaled radial distance where we introduced the radiation radius $R_0$. The scaling factors $N_w$ and $N_f$ are chosen such that the distributions are normalized, $\int w(r)\, dr=1$ and  $\int_0^\infty f(d_h,r_s) dd_h=1$.

It was observed that the optimum value for $R_0$ depends on the distance from $\Xmax$ to the shower impact point. 
For small distances there is an almost linear dependence until a saturation value of $R_0=50$~m is reached at a distance of about 5~km, independent of zenith angle.
A full account of the parametrization for fair weather showers in MGMR3D is presented in~\cite{Mitra:PhD}.

\subsection{Parameterization for thunderstorm showers}\seclab{Thunder}

Due to convection flows in clouds there is a build up of electric charge that has a layered structure when considering large clouds. In some cases this can spontaneously cause a lightning discharge. The charge layer structure depends on detailed cloud dynamics and temperatures, see for a recent reference~\cite{Salvador:2021}. To extract the atmospheric field configuration we thus parameterize a semi-realistic atmospheric field configuration as built in three different layers where in each layer the electric field may have a different orientation and strength. Including the boundary heights we thus arrive at nine parameters for the field in addition to two parameters for the location of the shower core as well as one for the energy of the cosmic ray and one for $\Xmax$. We assume that the arrival direction of the cosmic ray is known from the arrival time of the radio signals in the different antennas. Since this adds up to a large number of parameters it is imperative to limit the introduction of additional parameters that are not likely to greatly influence the structure of the extracted electric fields.
The values for the radiation radius $R_0$, the amount of charge excess, parameters defining the structure of the shower such as the first interaction depth $X_0$, and the width of the shower profile $\lambda$ are therefore kept fixed to the generic values used for fair-weather showers. The values of these parameters are given in Table.~\ref{table_values_paras} in the Appendix (except for $R_0$ that was discussed in the previous section).

There are two parameters that are more specific for the cases in which there is an electric field present. One is related to how quickly the induced transverse currents adjust to an electric field that changes with height, discussed in \secref{DES}. Another parameter is $u_0$ as introduced in \eqref{drift-v}. In \secref{single} the optimal values for these additional parameters are obtained by fitting an ensemble of showers simulated in CoREAS, following the approach introduced in ~\cite{Mitra:PhD}.

To verify the implementation of atmospheric electric fields in MGMR3D and to test the superposition principle we present in \secref{mult} a comparison with CoREAS calculations is performed for more intricate field configurations.

\subsubsection{Height-dependent electric fields}\seclab{DES}

In the presence of atmospheric electric fields the net force acting on the particles changes.
The component of the electric field parallel to the shower axis is assumed to be well below the runaway breakdown~\cite{Dwyer:2014}. This component can increase the number of secondary non-ultra-relativistic (with energies well below a GeV) electrons while decreasing the number positrons or the other way around, depending on the its orientation.
However, these particles are generally non relativistic and thus they trail far behind the shower front~\cite{Trinh:2016}.
For this reason, they do not contribute much to the radio radiation in the frequency range from 30~MHz to 80~MHz~\cite{Trinh:2016} and thus this field component can be ignored. 

We have noticed that electric fields in excess of 50~kV/m, that are transverse to the shower, do affect the structure of the shower profile.  We see an increase in the number of charged particles starting at the height where the electric field perpendicular to the shower axis, $\vec{E}_{\perp}$, is present. As a result, the value of $\Xmax$ is modified in the presence of a strong electric field. The reason for this effect is still unknown. To model this we modify the parametrization of the shower profile as
\bea
N_{c,E}(\Xrh)=N_{c}(\Xrh) \left(1 + 0.0015\,\frac{\rho(0)}{\rho(h)}\frac{F_{\perp}^2}{|v\times B|^2} \right)\;,
\eea
where $F_{\perp}$ is given in \eqref{F-perp}.

As has already been argued, $\vec{E}_{\perp}$, has a large effect on the radio emission. In the presence of atmospheric electric fields the transverse force acting on the shower particles is
\beq
\vec{F} = q(\vec{E}_{\perp} + \bf{v}\times\bf{B}) \;.\eqlab{F-perp}
\eeq
This changes both the magnitude and the direction of the drift velocity which is parallel to $\vec{F}$. In strong atmospheric electric fields, the net forces become large and thus the drift velocity increases. The simulations show that the radio emission does not increase proportionally and shows saturation effect for strong fields~\cite{Trinh:2016}. This can be understood from relativistic effects where the energy is no longer proportional to the square of the velocity. To account for this we follow Ref.~\cite{Trinh:2016} and introduce the saturation effects in the drift velocity, \eqref{drift-v}, by means of the parameter $u_0$.
In \secref{single}, \tabref{u0}, the value for this parameter is discussed.

Atmospheric electric fields may vary strongly with height. Thus another important parameter is the time, or distance, it takes for the drift velocity to adapt to the changes in the electric field. To parametrize this we replace $\vec{F}_\perp$ in \eqref{vh} by an effective force, $\vec{\tilde{F}}_\perp$, defined as
\bea
\vec{\tilde{F}}(h)&=&\vec{F}(h) + \nonumber \\
&&\sum_j \frac{1}{1+e^{(a+(h-H_s-h_j)\,b) }} e\Delta\vec{E}_j \;,
\eea
where
\beq
a = D_{Es}\,H_s\,\rho_{\rm ground}/X_{EM} \;,
\eeq
and
\beq
b=D_{Es}\,\rho(h_j)/X_{EM}\;.
\eeq
Here $\Delta\vec{E}_j$ is the change in the electric field at height $h_j$ and $X_{EM}$ is the interaction length for an electromagnetic particle. The parameter $D_{Es}$ (for E-smooth) governs the distance over which the drift velocity (through the effective force) adapts to the change in the electric field, if $D_{Es}$ is large the second form will reduce to a step function $\theta(h+H_s-h_j)$. The drift velocity will only adapt to the applied force after a certain relaxation time and the parameter $H_s=700$ is chosen such that the average shift in height, as shown later in \figref{height} is minimal.

\begin{figure}[H]
\centering
\includegraphics[width=0.5\textwidth]{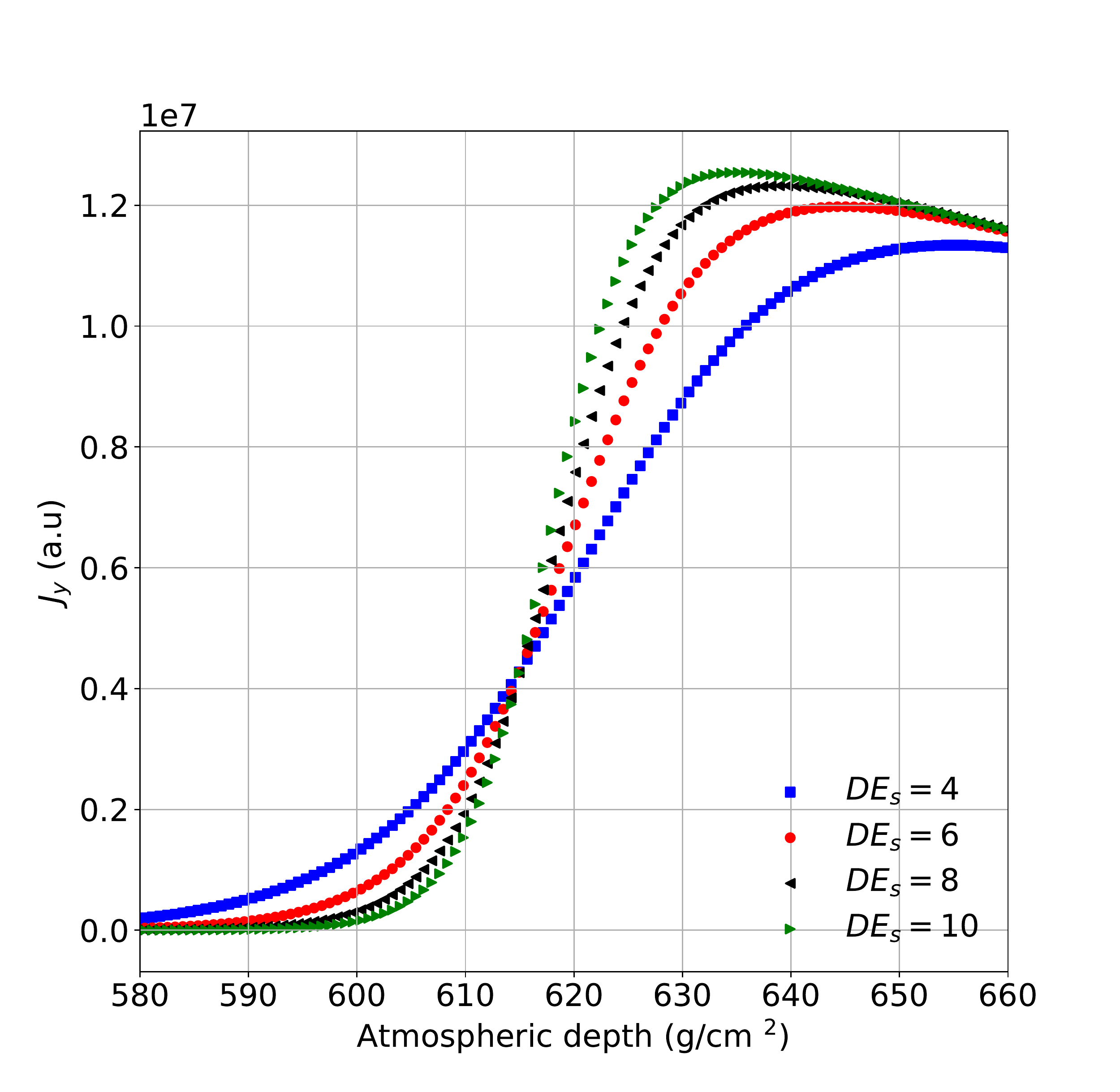}
\caption{The $\vvB$-component of the current profile for $D_{Es} = 4, 6, 8$ and $10$ as a function of atmospheric depth.}
\label{fig-DEs}
\end{figure}

Fig.~\ref{fig-DEs} shows how the current along the $\vvB$ direction changes when the EAS with $\Xmax=500$~g/cm$^2$ passes through an electric field boundary for different values of $D_{Es}$. An electric field of 50~kV/m is taken between a height of 4.5~km ($X=590$~g/cm$^2$) and the ground,  oriented along the $\vvB$-direction. It shows that the current changes faster when the value of $D_{Es}$ is larger. The drop-off of the current at large values of $X$, i.e.\ close to the ground, is due to the decrease in the number of shower particles.

To determine the optimal values for the  parameters $D_{Es}$ and $u_0$ we simulate a set of 94 vertical showers passing through a one-layer electric field using the microscopic code CoREAS. For each shower there is an electric field present from a certain height $h$ to the ground with certain strength $E$ and angle $\alpha$ where $\alpha$ is the angle between the electric field and $\vvB$-direction. The height $h$, the strength $E$, the angle $\alpha$, as well as $\Xmax$ are chosen randomly.

To select the optimal value for $D_{Es}$, we choose four different values for $D_{Es}$, 4, 6, 8, and 10.
As shown in Table.~\ref{DEs}, the mean value of $\chi^2$ does not change much when changing $D_{Es}$ and thus we do not need to determine $D_{Es}$ to higher precision.
While keeping this parameter fixed and using the height $h$, the strength $E$ and the angle $\alpha$ obtained from CoREAS, we perform a chi-square fit of $\Xmax$. We also keep $H_s$ fixed at 700 and $u_0 = 10$. By using a Levenberg-Marquardt minimization procedure based on a steepest descent method, we optimize the value of $\Xmax$ by minimizing 
\beq
\chi^2=\sum_{\text{antenna}}\sum_{S=I}^{Q,U,V}\left(\frac{S_{k,C}-f_rS_{k,3D}}{\sigma_k^2} \right )\,,
\eqlab{chi2}
\eeq
where $S_{k,C}$ are the Stokes parameters from CoREAS calculations for antenna at position $k$ and $S_{k,3D}$ are the ones from MGMR3D. $\sigma_k$ is the uncertainty as defined in~\eqref{err} and $f_r$ is the normalization factor for the radio intensity.
The results indicate that the value of $D_{Es}$ has only a minor effect on the fitted values of $\Xmax$. However, as shown in Table.~\ref{DEs}, the $\chi^2$ has a shallow minimum for  $D_{Es}=8$ . Therefore, we keep $D_{Es}$ fixed to this value for all subsequent calculations. The results of fitting also show that the value of $\chi^2$ does not depend on $\Xmax$ and the electric field configuration.

\begin{table}[]
\begin{tabular}{|c|c|c|c|c|}
\hline
\textbf{$D_{Es}$}                & 4     & 6     & 8     & 10    \\ \hline
\textbf{Mean $\chi^2$}      & 0.183 & 0.178 & 0.175 & 0.178 \\ \hline
\textbf{Standard deviation} & 0.156 & 0.150 & 0.146 & 0.145 \\ \hline
\end{tabular}
\caption{Mean $\chi2$ and its standard deviation for different values of $DEs$}
\label{DEs}
\end{table}

The other parameter which is kept fixed for all thunderstorm showers is $u_0$ as introduced in \eqref{drift-v}. Following the same procedure as  for  $D_{Es}$, fitting $\Xmax$ while keeping $u_0$, $h, E$ and $\alpha$ fixed, the mean $\chi^2$ and the standard deviation are calculated for several values of $u_0$ in the range from 0.3 to 10. We also see that the parameter $u_0$ hardly influences the extracted values of $\Xmax$. As shown in \tabref{u0}, the mean value of $\chi^2$ decreases when $u_0$ increases. However, when $u_0$ is equal or larger than 1, it does not affect the quality of the fit anymore, so we keep it fixed at the value of 10.
\begin{table}[]
\begin{tabular}{|c|c|c|c|c|c|c|}
\hline
\textbf{$u_0$}                & 0.3     & 0.6     & 1     & 3    & 7 & 10 \\ \hline
\textbf{Mean $\chi^2$}      & 0.188  & 0.177  & 0.176 & 0.176 &  0.175 & 0.176  \\ \hline
\textbf{Standard deviation} & 0.164 & 0.149 & 0.147 & 0.146 & 0.147 &  0.146\\ \hline
\end{tabular}
\caption{Mean $\chi2$ and its standard deviation for different values of $u_0$}
\tablab{u0}
\end{table}

\subsubsection{Accuracy of determined electric fields}\seclab{single}
We first investigate the accuracy of extracting one-layer electric fields.
For all 94 vertical showers passing through a one-layer electric field simulated by CoREAS as discussed in the previous section, we follow a fitting procedure using MGMR3D to reconstruct the structure of the field and compare this with the true values used in the CoREAS calculation. In this section the true electric field is taken homogeneous from a top height $h_{\rm true}$ to the ground with a strength $E_{\rm true}$, making an angle $\alpha_{\rm true}$ with the $\vvB$ axis. The reconstructed values, obtained by fitting the radio footprint using MGMR3D, carries a subscript ${\rm reco}$.

Since we have observed that fitting the field structure and $\Xmax$ at the same time sometimes results in ill-converging fit, we have taken the following approach.
We perform 20 reconstructions for fixed values for $\Xmax$ that vary in steps of 20~g/cm$^2$ between 500~g/cm$^2$ and 900~g/cm$^2$. For each construction, we fit the three parameters of the electric-field structure $h, E$ and $\alpha$ while keeping $\Xmax$ fixed. As discussed in Ref.~\cite{Trinh:2020}, from the Stokes parameters, we know the power value and the orientation of the polarization vector, however not the sign of the electric field vector. As a result, there are two solutions that give almost the same value of $\chi^2$ but differ in the sign of the induced current and thus that of the electric-field direction. For this reason, we change the obtained angle $\alpha$ by 180$^\circ$ and fit again for $h, E$ and $\alpha$ while keeping $\Xmax$ fixed. We select the electric field structure and the value of $\Xmax$ that give the smallest value for $\chi^2$.

\begin{figure}[h]
\centering
\includegraphics[width=0.5\textwidth]{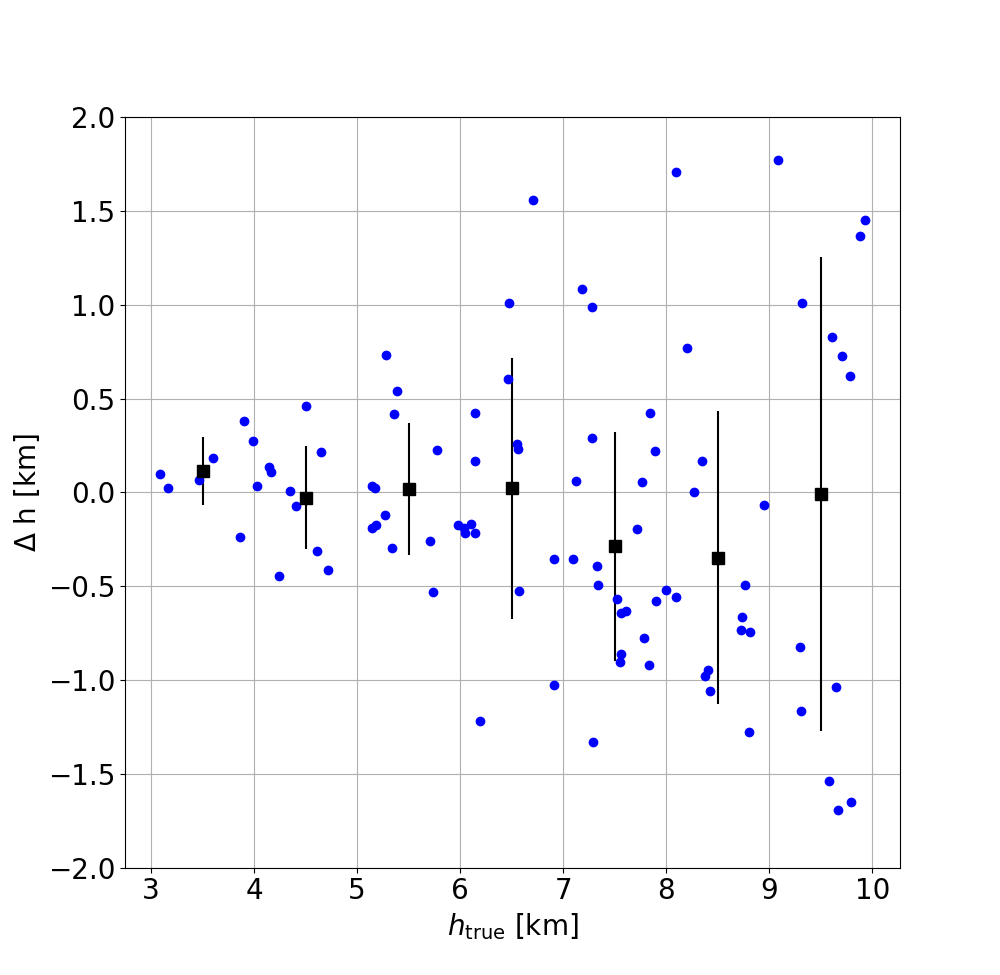}  
\caption{$\Delta h = h_{\rm reco} - h_{\rm true}$  as a function of the true height, $h_{\rm true}$. The black squares give the average value per kilometer bin while the error-bars denote the standard deviation.}
\figlab{height}
\end{figure}

In \figref{height} the difference  $\Delta h =  h_{\rm reco} - h_{\rm true}$ is shown as function of the true height where the electric field starts, $h_{\rm true}$, chosen randomly between 3~km and 10~km. Lower values were not considered as the agreement for these cases is close to perfect. 
Per height-bin of 1~km the mean values and the standard deviations are given. For heights lower than 9~km a rather good agreement is obtained between the reconstructed height, $h_{\rm reco}$, and the true height where the electric field starts, $h_{\rm true}$ with a standard deviation of less than 1~km. For heights between 9~km and 10~km, the spread in $\Delta h$ increases to reach a standard deviation of about 1.3~km although the mean difference is vanishingly small. The explanation is that at large altitudes there are not many particles, as the shower is still very young, and we thus loose sensitivity. 
Even though the spread in heights is increasing from 0.2~km at the height of 3.5~km to 1.3~km at 9.5~km it should be noted that the relative errors, $\Delta h/h_{\rm true}$, stay more constant, ranging from 5\% to 13\%, in a similar range as for the single layer case.

\begin{figure}[h]
\centering
\includegraphics[width=0.5\textwidth]{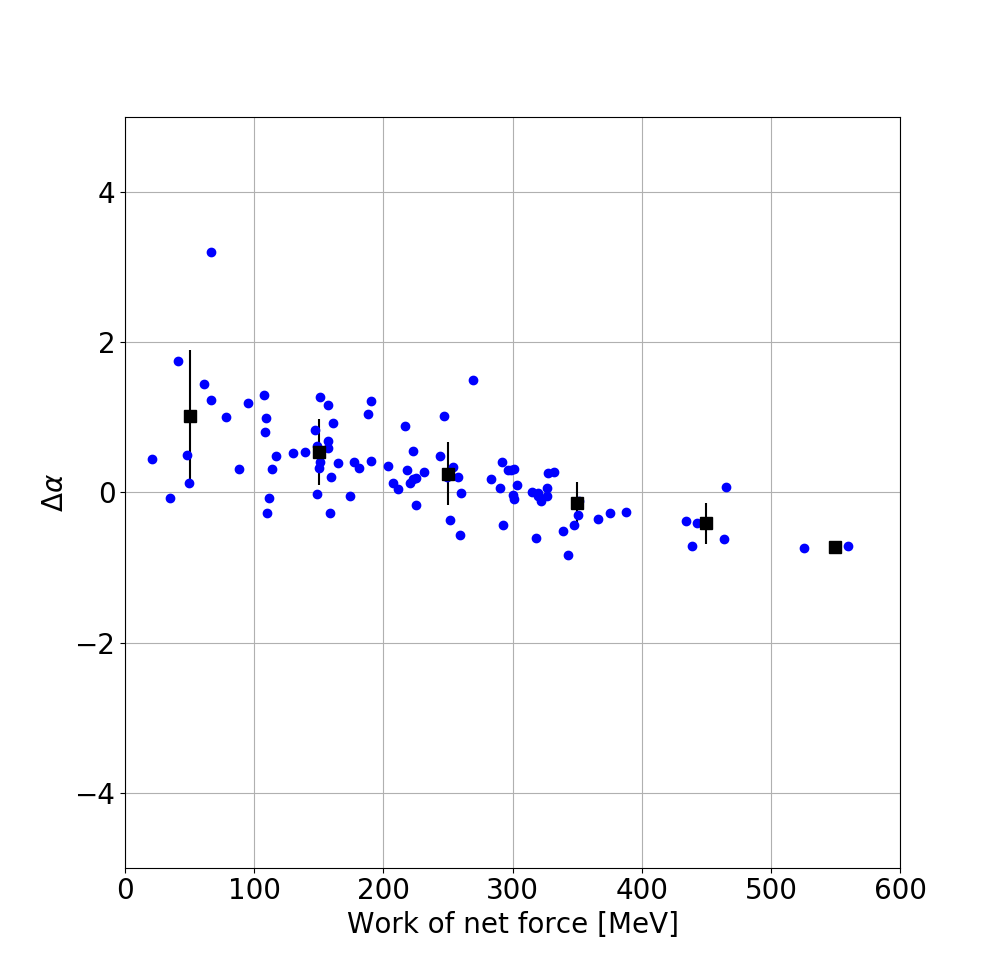}   
\caption{$\Delta \alpha = \alpha_{\rm reco} - \alpha_{\rm true}$  as a function of the work of the net force $W$. The black squares give the average value per 100-MeV bin while the error-bars denote the standard deviation.}
\figlab{alpha}
\end{figure}

The true value of the electric field $E_{\rm true}$ is also chosen randomly between 5~kV/m and 70~kV/m while the angle $\alpha$ is randomly selected from 0 to 360$^\circ$. 
In \figref{alpha} we plot $\Delta \alpha = \alpha_{\rm reco} - \alpha_{\rm true}$ as a function of the work of the net force $W=F.H$. Here $F$ is the net force of the Lorentz force and the true electric force and $H$ is the thickness of the layer. For these one-layer cases, $H=h_{\rm true}$. When the work of the net force of a layer is large, the amount of radio emission emitted from this layer is also large and thus we are more sensitive to this layer. As a result, the orientation of the electric field in this layer is determined more accurately (or $\Delta\alpha$ is small) for large values of $W$. \figref{alpha} shows that the angle of the field is very well reconstructed, with $\Delta \alpha = \alpha_{\rm reco} - \alpha_{\rm true} < 4^\circ$. For work less than 100~MeV, the deviation of $\Delta \alpha$ is about 1$^\circ$, while for larger work of the net force, it almost vanishes.

\begin{figure}[h]
\centering
\includegraphics[width=0.5\textwidth]{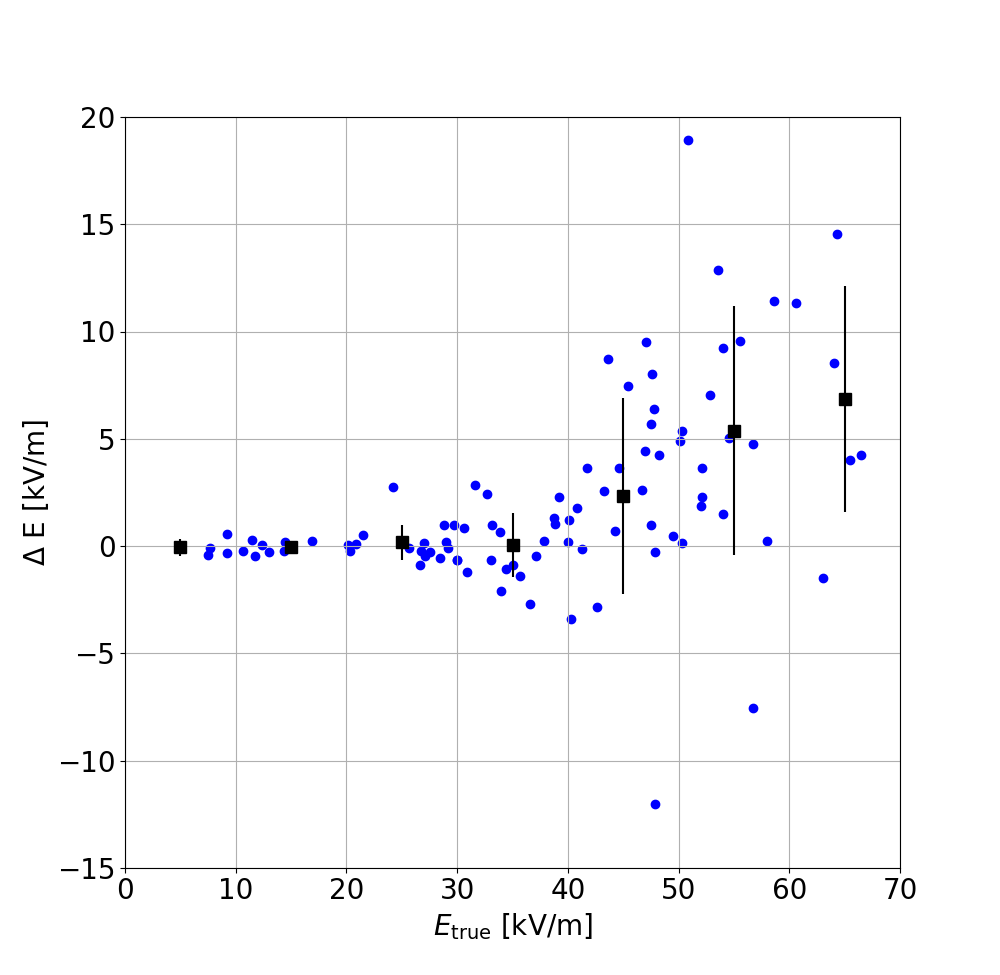}  
\caption{$\Delta E= E_{\rm reco} - E_{\rm true}$ as a function of the true strength of the electric field, $E_{\rm true}$. The black squares give the average value per 10-kV/m bin while the error-bars denote the standard deviation.}
\figlab{E}
\end{figure}

\figref{E} shows the difference between the reconstructed strength of the electric field and the true value, $\Delta E =E_{\rm reco}-E_{\rm true}$.
When the electric field is less than 40~kV/m, the reconstructed values from the MGMR3D calculation agrees well with the true value, with differences less than 5~kV/m. When the field strength is large, the differences become large since the current saturates at about 50~kV/m and thus we loose the sensitivity to the field strength.
However, the relative errors are less than 5\% when the field strenght is weaker than 50~kV/m and vary from 8\% to 11\% when the field strength is large.

\begin{figure}[h]
\centering
\includegraphics[width=0.5\textwidth]{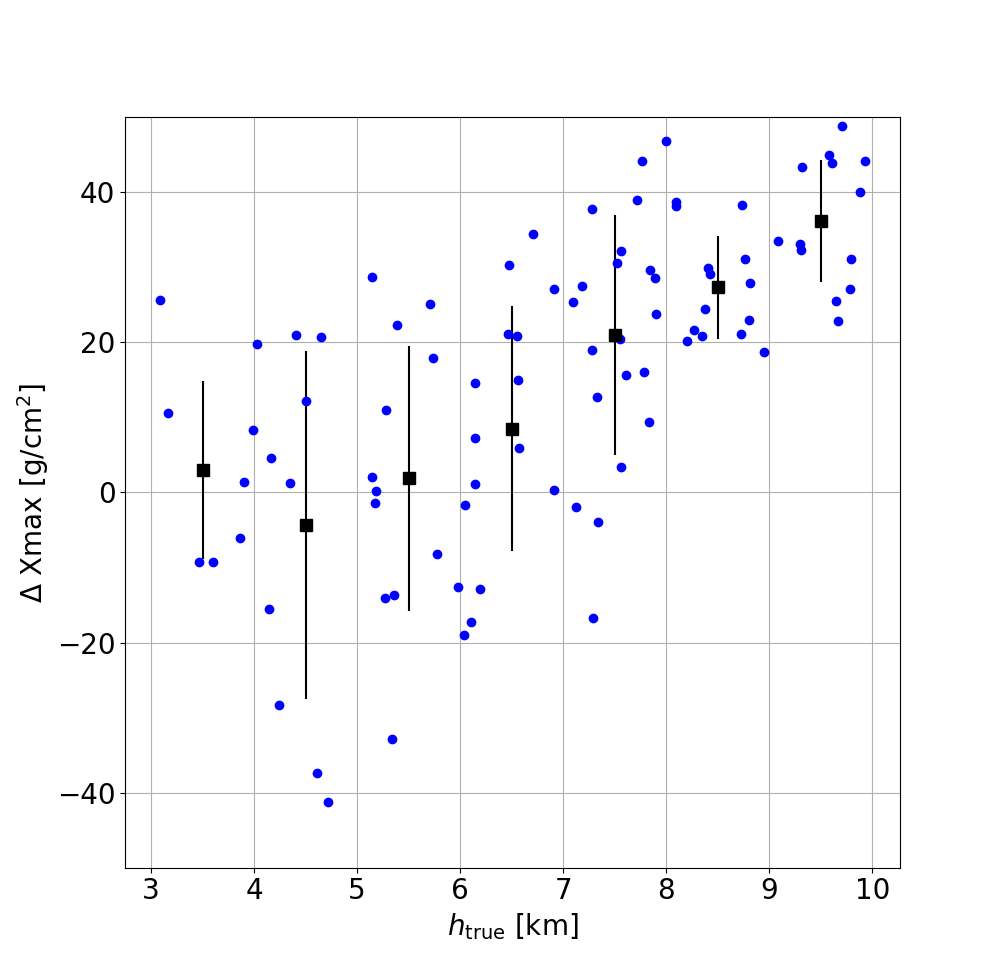}  
\caption{$\Delta \Xmax=\Xmax$$_{\rm reco}$$ - \Xmax$$_{\rm true}$ as a function of the true height $h_{\rm true}$. The black squares give the average value per kilometer bin while the error-bars denote the standard deviation.}
\figlab{Xmax}
\end{figure}

\figref{Xmax} shows the difference between the reconstructed values of $\Xmax$ and the true values, $\Delta \Xmax=\Xmax$$_{\rm reco}$$ - \Xmax$$_{\rm true}$.
For heights lower than 7~km, there seems to be no sizable systematic differences. The mean discrepancies are less than 10~g/cm$^2$, comparable to the fair-weather results for $\Xmax$.
For heights from 7~km to 10~km, the mean difference is a bit larger, from 30 ~g/cm$^2$ to 40~g/cm$^2$, and increases with the true height, but the standard deviation is smaller.

One other important parameter is the normalization factor $f_{r}$ for the Stokes parameters between MGMR3D and CoREAS which is given in \eqref{chi2}.
For fair-weather showers, the mean normalization factor is equal to 1. For thunderstorm showers, the mean value of the normalization factor of these 94 showers is 0.63 and the standard deviation is 0.24. This means that the radiation in MGMR3D is overestimated as compared to the CoREAS result. This could be a reflection of the fact that the reconstructed electric fields are a bit over estimated, as shown in \figref{E}.

As was shown in this section, there are no sizable systematic differences between the reconstructed parameters for the electric field structure and $\Xmax$ using a MGMR3D fit and the Monte-Carlo truth as expressed using CoREAS. The systematic differences are mostly smaller than the reconstruction accuracy.

\subsubsection{Multiple layers}\seclab{mult}

To obtain test results for semi-realistic cases we have simulated showers passing through a two-layer electric field.  Layer 1 extents  from height $h_1$ to the height $h_2<h_1$. Layer 2 extents from height $h_2$ to the ground.
The height for layer 1, $h_1$, is selected in the range from 5~km to 10~km. The height for layer 2, $h_2$, is chosen in the range from 2~km to 4.5~km. In reality the field strength in layer 1 is usually larger than the one in layer 2. Thus, the field strength in layer 1, $E_1$, and layer 2, $E_2$, are chosen randomly from 2~kV/m to 60~kV/m and from 2~kV/m to 30~kV/m, respectively, each with arbitrary angles $\alpha_1$ and $\alpha_2$ with respect to the $\vvB$-axis.
The field strength $E_2$ is chosen smaller than $E_1$ since in reality the electric field in the cloud is usually stronger than the one between the cloud to ground.
Within these constraints the parameters are chosen randomly to simulate 91 vertical showers in CoREAS.
In order to find the reconstructed values using MGMR3D, following a similar procedure as was taken for the one-layer case, we perform 20 separate parameter searches with different values of $\Xmax$. We fit the six parameters of the two-layer electric field while keeping $\Xmax$ fixed. We then repeat the procedure changing the values of $\alpha$ by 180$^\circ$.

\begin{figure}[h]
\centering
\includegraphics[width=0.5\textwidth]{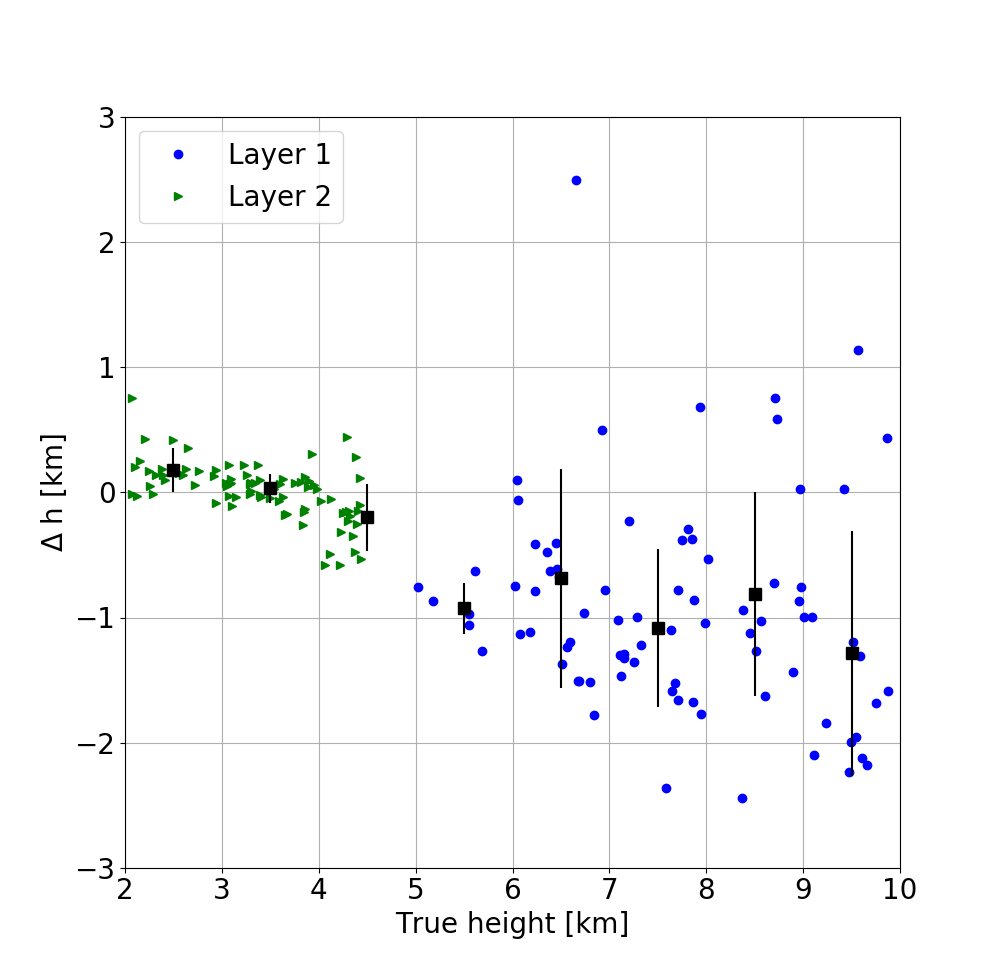} 
\caption{$\Delta h = h_{\rm reco} - h_{\rm true}$ as a function of the true height, $h_{\rm true}$, for layer 1 and layer 2. The black squares give the average value per kilometer bin while the error-bars denote the standard deviation.}
\figlab{height21}
\end{figure}

\figref{height21} shows the difference between the reconstructed and the true height, $\Delta h = h_{\rm reco}- h_{\rm true}$, for both layers. We observe very similar differences, of the same order of magnitude as shown in \figref{height} for the single-layer case. For small heights in layer 2 the standard deviation is of the order of 100~m, but is larger, of the order of 1000~m, for larger heights in layer 1. On average the top height is reconstructed at a too low altitude. Similar to what is seen in the single-layer case, although the spread in heights for layer 1 is much larger than that for layer 2, the relative errors vary less, from 5\% to 13\%.

\begin{figure}[h]
\centering
\includegraphics[width=0.5\textwidth]{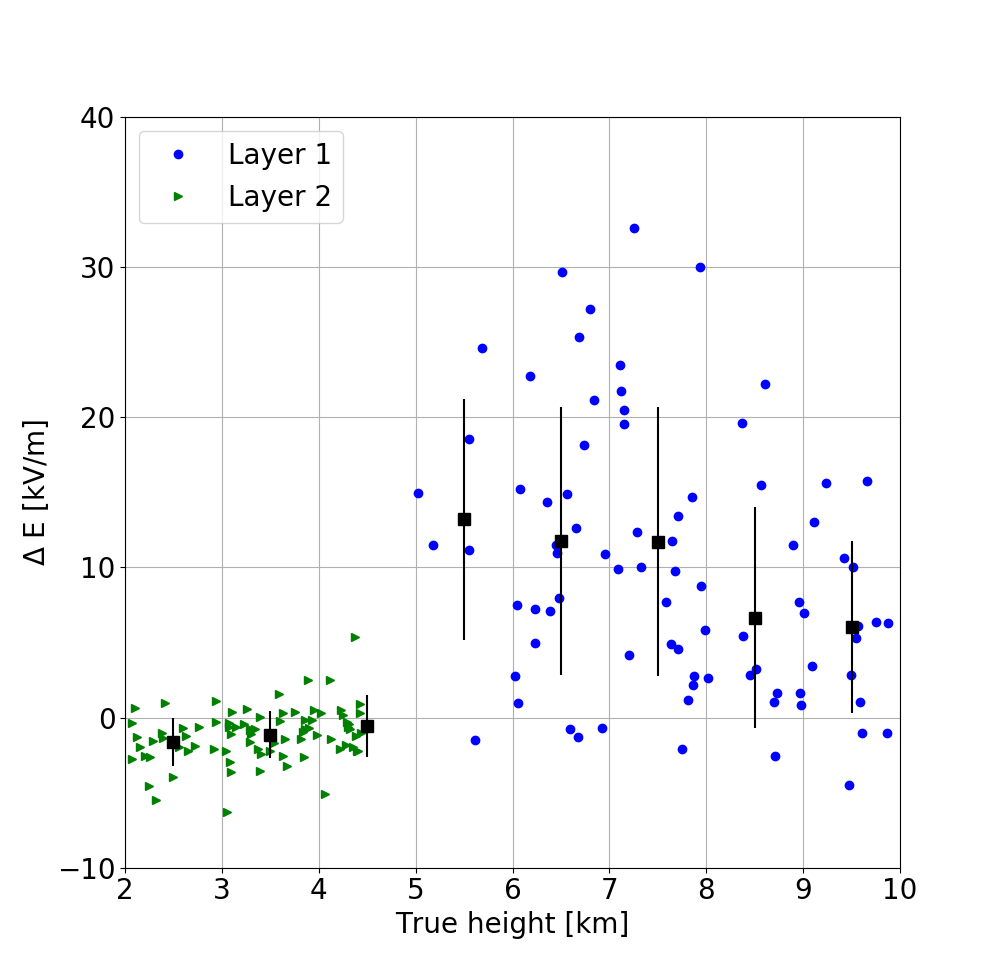}  
\caption{$\Delta E = E_{\rm reco} - E_{\rm true}$ as a function of the true height, $h_{\rm true}$, for layer 1 and layer 2. The black squares give the average value per $10-kV/m$ bin while the error-bars denote the standard deviation.}
\figlab{E21h}
\end{figure}

Since we have found that the difference between reconstructed and true electric field strength correlates most strongly with the upper height of the layer, we show in \figref{E21h} $\Delta E= E_{\rm reco}- E_{\rm true}$ as function of the height. We observe that the standard deviation of $\Delta E$ differs for the two layers. While the mean $\Delta E$ is constant and slightly negative for the lower layer, the mean value for $\Delta E$ is positive and decreases with height for the upper layer. The spreads in layer 1 are larger than that in layer 2 because the field strength in layer 1 is stronger than the one in layer 2 and for large field strength we loose sensitivity as shown in the one-layer case.

\begin{figure}[h]
\centering
\includegraphics[width=0.5\textwidth]{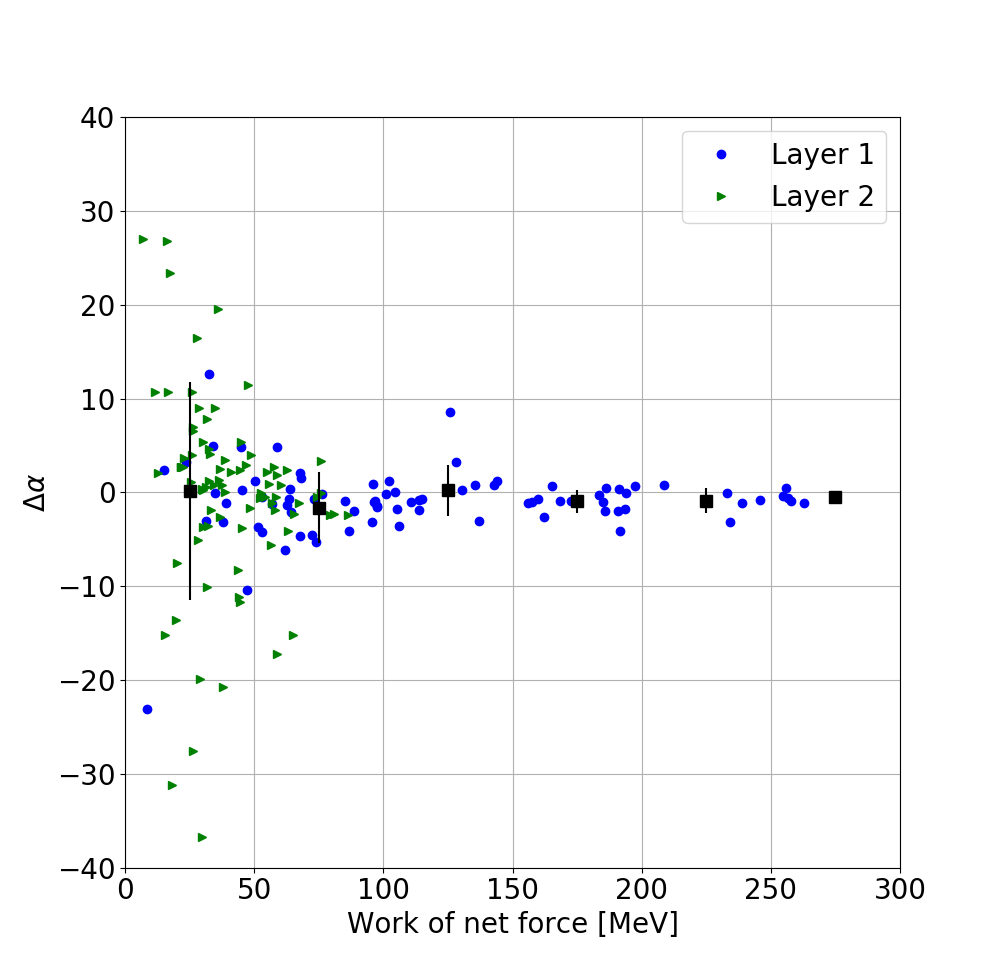} 
\caption{$\Delta \alpha = \alpha_{\rm reco} - \alpha_{\rm true}$ as a function of the true height, $h_{\rm true}$, for layer 1 and layer 2. The black squares give the average value per 50-MeV bin while the error-bars denote the standard deviation.}
\figlab{alpha21}
\end{figure}

\figref{alpha21} shows that there is a correlation between $\Delta \alpha$ and the work of the net force $W$. When the work of the net force of a layer is larger than 50~MeV, which is usually seen in layer 1, the amount of radio emission emitted from this layer is large and thus we are quite sensitive to this layer. Therefore, the orientation of the electric field in this layer is determined rather accurately or $\Delta \alpha$ is small. In contrast, when the work of the net force is smaller than 50~MeV, which is usually seen in layer 2, the amount of radiation from this layer is small and thus we loose sensitivity to the polarization or the spread in $\Delta \alpha$ is large.

\begin{figure}[h]
\centering
\includegraphics[width=0.5\textwidth]{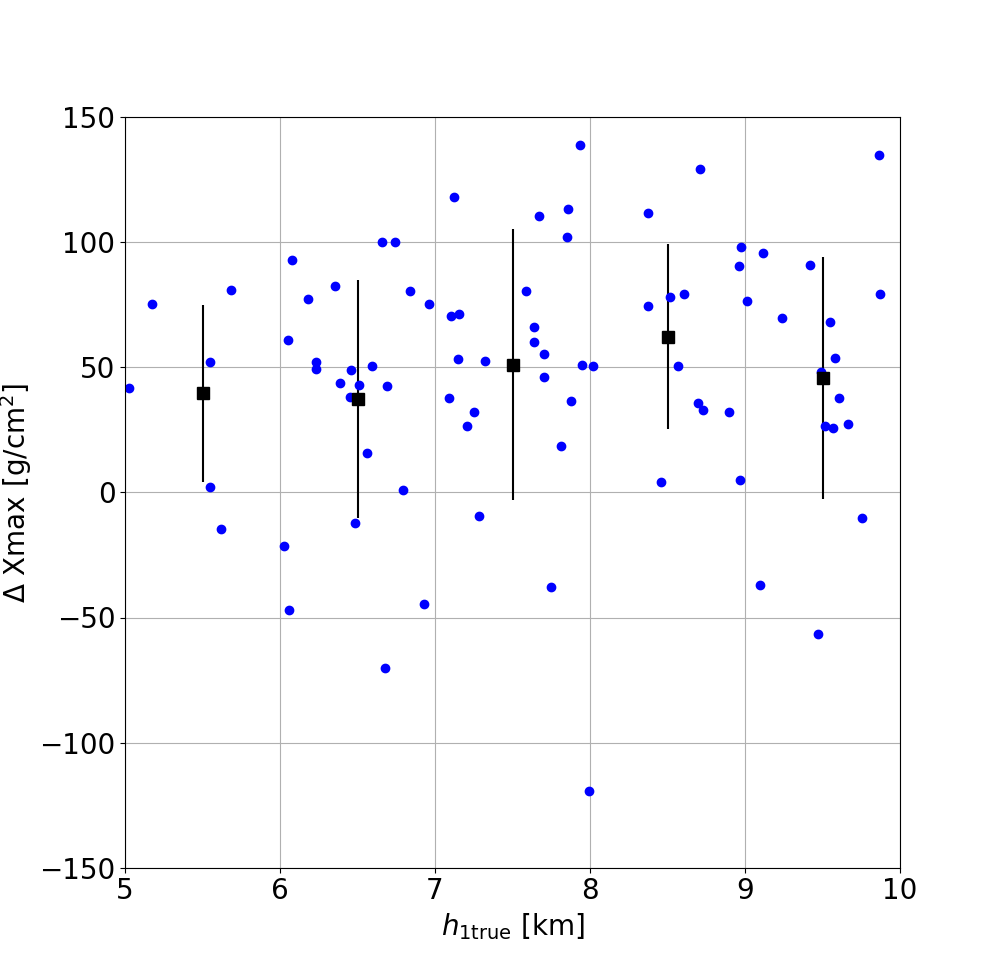}  
\caption{$\Delta \Xmax = \Xmax$$_{\rm reco}$$-\Xmax$$_{\rm true}$  as a function of the true height of the top layer $h_{1\rm true}$. The black squares give the average value per kilometer bin while the error-bars denote the standard deviation.}
\figlab{Xmax21}
\end{figure}

\figref{Xmax21} shows that the differences for $\Xmax$ are on average positive with a mean difference of about 50~g/cm$^2$, about equal to the standard deviation. The fact that $\Xmax$ is deeper in the atmosphere is probably correlated with the, on average, negative values for $\Delta h$ shown in \figref{height21}.

Similar to the one-layer showers, the mean norm factor and its standard deviation of the two-layer showers are $0.62$ and $0.19$, respectively.

To conclude, in general the reconstructed values are scattered around the true values and the strongest bias is seen in the obtained values for $\Xmax$ that are, however, of little interest in the study of atmospheric electric fields.

\section{ Extracted electric field configuration from LOFAR data using MGMR3D}\seclab{appl}

To test the proposed approach for reconstructing the atmospheric electric field from data, we have applied the reconstruction procedure to LOFAR data for a cosmic ray event that has been recorded under thunderstorm conditions~\cite{Trinh:2020}.
We selected an event for which the radio footprint is particularly complicated, to the extent we could not reach satisfactory results in an earlier application of the procedure,  event~1 from Ref.~\cite{Trinh:2020}. The shower axis for  this event is inclined with a zenith angle of $\theta = 39.2^{\circ}$. As shown in Ref.~\cite{Trinh:2020}, there were large differences between CoREAS and the reconstruction for this event using the older version of MGMR3D.

\begin{figure*}
\centering
\includegraphics[width=\textwidth]{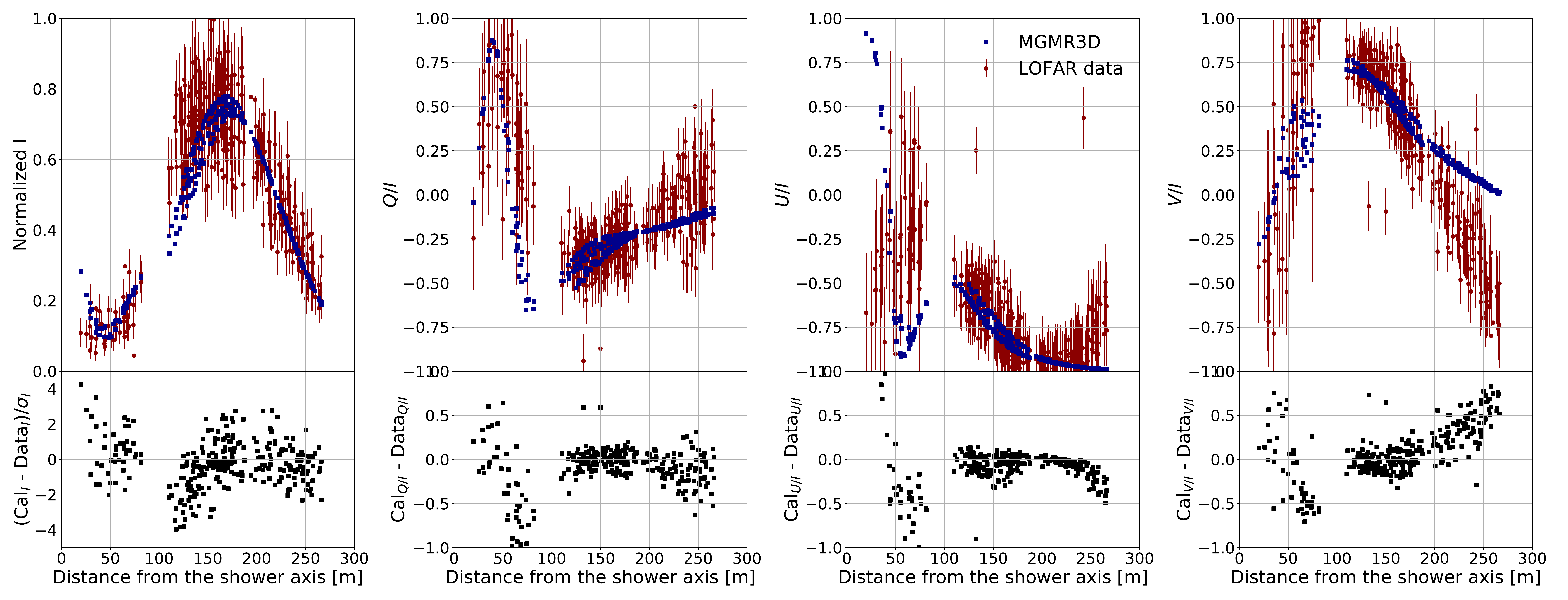}
\caption{The results of MGMR3D calculation for Stokes parameters (filled blue dots) are compared to LOFAR data (filled red circles) for event~1. $\sigma$ denotes one standard deviation error.}
\label{MGMR}
\end{figure*}

\begin{figure*}
\centering
\includegraphics[width=\textwidth]{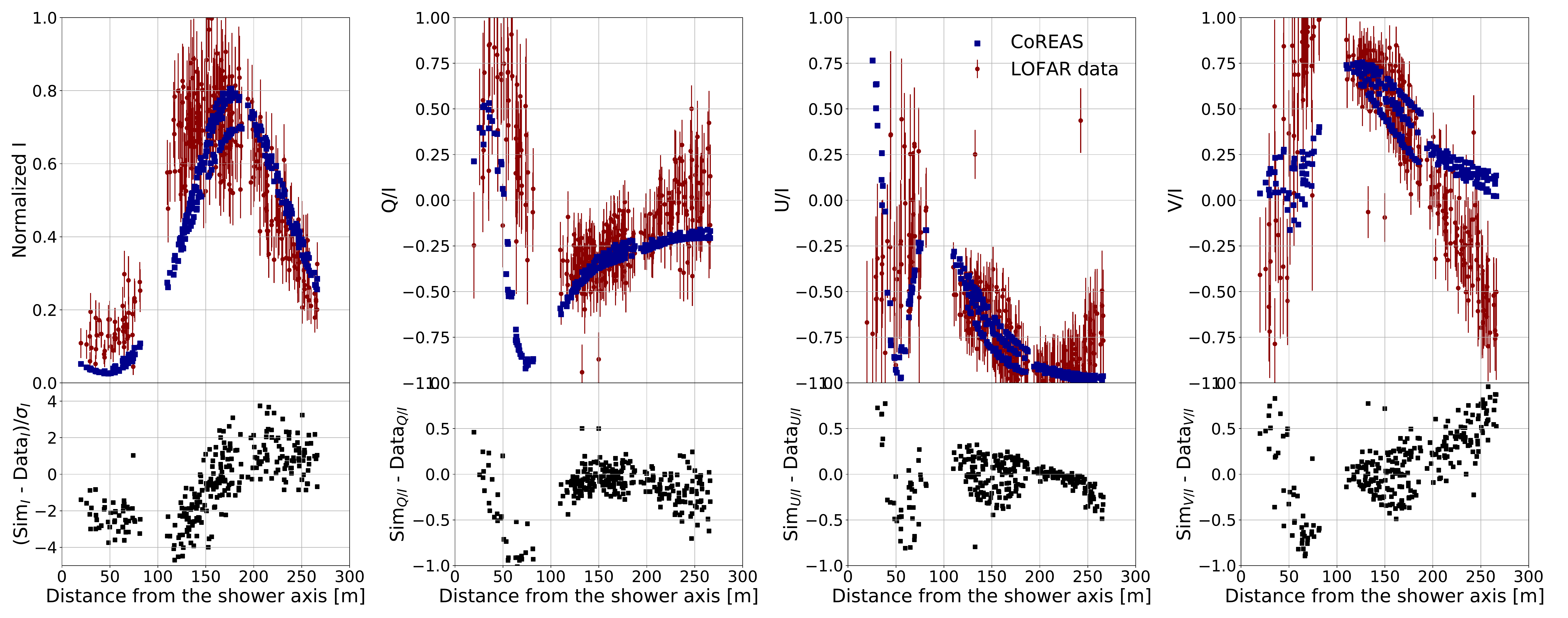}
\caption{Same as Fig.~\ref{MGMR} but for the comparison between CoREAS simulation and LOFAR data for event~1}
\label{CoREAS}
\end{figure*}

We have applied the present reconstruction method to recover the electric field for this event. Similar to what has been done in Ref.~\cite{Trinh:2020}, we model the electric field with three layers. The core position is kept fixed at the position estimated from the particle-detector array LORA~\cite{Thoudam:2014}. We perform 20 reconstructions for fixed values for $\Xmax$ that vary in steps of 20~g/cm$^2$ between 500~g/cm$^2$ and 900~g/cm$^2$. For each construction, we fit nine parameters of the electric-field structure while keeping $\Xmax$ fixed. At the end we select the electric field structure and the value of $\Xmax$ that give the best fit result. This approach is similar to the reconstruction procedure used in~\secref{single} and~\secref{mult}. The results of the MGMR3D calculation that gives the best fit is shown in Fig.~\ref{MGMR}. The electric field reconstructed from this calculation is plugged in CoREAS to confirm the procedure.
Since in CoREAS, $\Xmax$ is an output which cannot be chosen before running simulations, we simulate 20 showers with the reconstructed electric-field structure and select the simulation which gives the smallest value of $\chi^2$, shown in Fig.~\ref{CoREAS}.
The values of the extracted electric field parameters, the value of $\Xmax$, the reconstructed energy and the values of $\chi^2$ are given in Table.~\ref{event1}.

With the new version of MGMR3D, the event can be reconstructed rather well by a three-layer electric field. As shown in Fig.~\ref{MGMR} and Fig.~\ref{CoREAS}, the differences between MGMR3D and CoREAS are relatively small. It can be seen from the first panels of these two figures that Stokes I, i.e. the radiation intensity, given by both MGMR3D and CoREAS fits the LOFAR data well, except for distances near the core of the shower. At these close distances, the intensity given by CoREAS is a bit lower than the data, with the peak shifted to a slightly larger distance. The intensity at small distances is rather small because of a destructive interference between the radiation from layer 2 and layer 3. The peak in the Stokes I at a distance of 175~m from the shower axis is reproduced well in the MGMR3D calculation.
Stokes $Q$ and Stokes $U$ represent the linear polarization of the radiation. The second and the third panels of Fig.~\ref{MGMR} and Fig.~\ref{CoREAS} show the ratio $Q/I$ and $U/I$, respectively. The polarization of the radiation is complicated, changing with the distances from the shower axis.
Stoke $V$, presenting the circular polarization, is plotted in the last panels of Fig.~\ref{MGMR} and Fig.~\ref{CoREAS} in terms of the ratio $V/I$. There is a large amount of circular polarization which is reflected in the change in the direction of the electric field between three layers.
The linear polarization and the circular polarization are also reproduced quite well except at small distances, within 100~m from the core of the shower where the radio intensity is rather small.
As shown in Table.~\ref{event1} the MGMR3D and CoREAS results correspond to an energy of the cosmic ray of $1\cdot10^8$~GeV and $1.07\cdot10^8$~GeV. This energy differs from what is deduced from the particle detectors, $2.67\cdot10^{7}$~GeV. The reason for this could be that the response of the LORA detectors is affected by the thunderstorm. 

Since we can determine the components of the electric fields perpendicular to the direction of the shower, we can derive the purely horizontal components of the field. We decompose $\textbf{E}_\perp$ into two components $\textbf{E}_\textbf{vxz}$ and $\textbf{E}_\textbf{vx(vxz)}$ along $\textbf{vxz}$ and $\textbf{vxz}$ and shown in Table.~\ref{Ehor_event1}. The purely horizontal components $\textbf{E}_\textbf{vxz}$ determined for this event are small because this shower is inclined, with the zenith angle of $\theta = 39.2^{\circ}$, since one expects  the atmospheric electric field to have a strong vertical component. This component in layer 1 is larger than that in other lower layers, which is one would expect because the charge layers are not purely horizontal or the event occurred at the edge of the charged layer. The heights where the electric fields change could be the positions of different charged layers. There could be a negative-charge layer at 3.38~km and positive-charge layers at 8~km and 1.85~km altitude. This is similar to what has been determined from Lightning Mapping Array (LMA) observations of a flash that occurred near the LOFAR core in June 2019. Details about the interpretation of this flash can be found in Ref.~\cite{Trinh:2020}.

The result of performing a reconstruction for this complicated-footprint shows that we can get a stable result from the new version of MGMR3D that captures the main structures seen in intensity and polarization. Using the reconstructed fields in the microscopic calculation gives results for the complicated radio footprint that show the same, very non-trivial, structures.

\begin{table}[]
\begin{tabular}{|c|c|c|c|}
\hline
\textbf{Layer} & \textbf{h {[}km{]}} & \textbf{E {[}kV/m{]}} & \textbf{$\alpha$ {[}$^\circ${]}} \\ \hline
1      & 8.012       & 34.900        & 157.3                    \\ \hline
2      & 3.381       & 99.377        & -64.6                    \\ \hline
3      & 1.851       & 71.756        & 102.2                    \\ \hline
\multicolumn{3}{|c|}{$\Xmax$$_{\rm 3D}${[}g/cm$^2${]}} & \multicolumn{1}{c|}{860} \\ \hline
\multicolumn{3}{|c|}{$\Xmax$$_{\rm C}${[}g/cm$^2${]}} & \multicolumn{1}{c|}{848} \\ \hline
\multicolumn{3}{|c|}{Energy$_{\rm 3D}$[GeV]} & \multicolumn{1}{c|}{$1.0\times 10^8$} \\ \hline
\multicolumn{3}{|c|}{Energy$_{\rm C}$[GeV]} & \multicolumn{1}{c|}{$1.07\times 10^8$} \\ \hline
\multicolumn{3}{|c|}{$\chi^2_{\rm 3D}$} & \multicolumn{1}{c|}{$2.39$} \\ \hline
\multicolumn{3}{|c|}{$\chi^2_{\rm C}$} & \multicolumn{1}{c|}{$3.76$} \\ \hline

\end{tabular}
\caption{Nine extracted electric field parameters, $\Xmax$ and the reconstructed energy of event~1}
\label{event1}
\end{table}

\begin{table}[h]
\begin{tabular}{|c|c|c|c|}
\hline
\textbf{Layer} & \textbf{h [km]} & $\textbf{E}_\textbf{vxz}$ \textbf{[kV/m]} & $\textbf{E}_\textbf{vx(vxz)}$ \textbf{[kV/m]}\\ \hline
1      & 8.012       & 24.721        & 24.635                    \\ \hline
2      & 3.381       & -5.547        & -99.222                    \\ \hline
3      & 1.851       & -12.460        & 70.665                    \\ \hline
\end{tabular}
\caption{The components $\textbf{E}_\textbf{vxz}$ and $\textbf{E}_\textbf{vx(vxz)}$ of the electric fields determined from event~1}
\label{Ehor_event1}
\end{table}

\section{Summary and Conclusions}
In this work, we have made an comprehensive comparison between the radio emission during thunderstorm conditions from MGMR3D and from CoREAS. As a result, we improve the parameterization for MGMR3D in the presence of atmospheric electric fields.

With the parameters determined, we have generated a large number of radio footprints for showers passing through a layered electric field using the microscopic code CoREAS. From these footprints we have reconstructed the electric field structure using MGMR3D and compared these with the true structure that was used for generating the footprint. This shows that the layer heights can be reconstructed with an accuracy of better that $\pm 1$~km, the field strengths with $\pm 10$~kV/m, and orientation angles within a few degrees. Exceptions are the orientation of weak fields for cases in which there are also strong fields involved as well as cases where the top height of the electric field layer lies at a height where the shower is still young with relatively few charged particles.
In the region where we are sensitive to the parameters of the electric fields, the differences between MGMR3D and CoREAS are rather small for all vertical showers passing through a one-layer electric field or a two-layer one.
Large discrepancies are observed for the cases where we loose sensitivity to the parameters of the electric fields seen in both MGMR3D and CoREAS.

We have applied the approach to extract electric field structure to an event having a complicated radio footprint as measured by LOFAR and we are able to reconstruct the main features of this event.
Therefore, it can be concluded that MGMR3D can be used to reconstruct the structure of the electric fields by using the radio emission emitted from air showers passing through thunderclouds. This method of determining atmospheric electric fields will help to study about the process of lightning initiation and propagation.
\section*{Acknowledgement}
This research is funded by Vietnam National Foundation for Science and Technology Development (NAFOSTED) under grant number 103.01-2019.378. K. D. de Vries was supported by the European Research Council under the EU-ropean Unions Horizon 2020 research and innovation programme (KdV, grant agreement No 805486).

\appendix

\section{Parameter values}
\begin{table}[h]
\begin{tabular}{|l|c|c|c|c}
\cline{1-4}
 Parameter & $J_Q$ & $X_0$ & $\lambda$ &  \\ \cline{1-4}
 Value &  0.21 & 100 & 100 &  \\ \cline{1-4}
\end{tabular}
\caption{Fixed values of parameters}
\label{table_values_paras}
\end{table}
\section{Programming details}

The program can be downloaded from \href{https://drive.google.com/file/d/1ojpdk1B-5Iv0t7XCDaQGAUjQj-vqN7uP/view?usp=sharing}{MGMR3D-v3} as a zip file. Make sure you run version 3.

\bibliography{thunderstorm}
\end{document}